\documentclass[prb,
showpacs,
twocolumn,
floats,
10pt,
aps,
citeautoscript,
longbibliography,
superscriptaddress]{revtex4-2}

\usepackage{placeins}
\usepackage{lipsum}
\usepackage{xcolor}
\usepackage[normalem]{ulem}
\usepackage{comment}
\usepackage{tabularx}
\usepackage{graphicx}
\usepackage{dcolumn}
\usepackage{bm}
\usepackage{ulem}

\usepackage{bbm}
\usepackage{blindtext}
\usepackage{graphics}
\usepackage{verbatim}   
\usepackage{amsfonts}
\usepackage{amsmath}
\usepackage{amssymb}
\usepackage{adjustbox}
\usepackage{microtype} 
\allowdisplaybreaks 
\usepackage{xspace} 
\usepackage{multirow} 
\usepackage{tabstackengine}
\setstackEOL{\cr}

\usepackage{lmodern}
\usepackage[T1]{fontenc}

\newcommand{\Eq}[1]{Eq.~\eqref{#1}}

\newcommand{\Fig}[1]{Fig.~\ref{#1}}

\newcommand{\Sec}[1]{Sec.~\ref{#1}}
\newcommand{\App}[1]{App.~\ref{#1}}

\newcommand*{\ndots}{\kern-0.075em.\kern-0.05em.\kern-0.05em.}  
\newcommand*{\nidots}{.\kern-0.05em.\kern-0.05em.} 
\newcommand*{\ncdots}{\kern-0.15em\cdot\kern-0.2em\cdot\kern-0.2em\cdot\kern-0.15em}   

\NewDocumentCommand{\doubleI}{O{}}{\mathbbm{1}_{#1}}
\NewDocumentCommand{\doubleIb}{O{}}{{\overline{\mathbbm{1}}_{#1}}}
\NewDocumentCommand{\doubleIk}{O{}}{\mathbbm{1}^\ks_{\! #1}}
\NewDocumentCommand{\doubleId}{O{}}{\mathbbm{1}^\ds_{\! #1}}
\NewDocumentCommand{\doubleIp}{O{}}{\mathbbm{1}^\ps_{\! #1}}
\NewDocumentCommand{\doubleV}{O{}}{\mathbbm{V}_{\! #1}}
\NewDocumentCommand{\doubleVk}{O{}}{\mathbbm{V}^\ks_{\! #1}}
\NewDocumentCommand{\doubleVd}{O{}}{\mathbbm{V}^\ds_{\! #1}}
\NewDocumentCommand{\doubleVp}{O{}}{\mathbbm{V}^\ps_{\! #1}}
\NewDocumentCommand{\doublev}{o}{{\mathbbm{v}_{#1}}}
\NewDocumentCommand{\doubleVb}{o}{{\overline{\mathbbm{V}}_{\! #1}}}
\NewDocumentCommand{\doubleVt}{o}{{\widetilde{\mathbbm{V}}_{\! #1}}}
\NewDocumentCommand{\doubleVh}{o}{\widehat{{\mathbbm{V}}_{\! #1}}}
\NewDocumentCommand{\doubleW}{o}{\mathbbm{W}_{\! #1}}
\NewDocumentCommand{\doubleWk}{o}{\mathbbm{W}^\ks_{\! #1}}
\NewDocumentCommand{\doubleWd}{o}{\mathbbm{W}^\ds_{\! #1}}
\NewDocumentCommand{\doubleWb}{o}{{\overline{\mathbbm{W}}_{\! #1}}}
\NewDocumentCommand{\doubleWt}{o}{{\widetilde{\mathbbm{V}}_{\! #1}}}
\NewDocumentCommand{\doubleWh}{o}{{\widehat{\mathbbm{V}}_{\! #1}}}

\newcommand{\LMUMunich}{Department of Physics and Arnold Sommerfeld Center for Theoretical Physics (ASC), Ludwig-Maximilians-University Munich,
Theresienstr. 37, D-80333 Munich, Germany}

\newcommand{\MCQST}{Munich Center for Quantum Science and Technology, Schellingstr. 4, D-80799 Munich, Germany}

\newcommand{\Regensburg}{Institut für Theoretische Physik, Universität Regensburg, D-93035 Regensburg, Germany}

\definecolor{darkgreen}{rgb}{0,0.5,0}
\definecolor{purple}{rgb}{0.6,0,0.5}
\definecolor{orange}{rgb}{1,0.5,0}
\definecolor{darkred}{rgb}{.7,0,0}
\definecolor{darkblue}{rgb}{0,0,.6}
\definecolor{grey}{rgb}{.6,.6,.6}
\definecolor{dimgreen}{rgb}{0.2,0.7,0.2}
\definecolor{brightgreen}{rgb}{0.5020, 1, 0}

\newcommand{\jvdomit}[1]{}

\usepackage{multirow}
\usepackage{physics}
\usepackage{hyperref}
\hypersetup{colorlinks=true,breaklinks,linkcolor=blue,urlcolor=blue,citecolor=blue}

\usepackage{graphicx}
\usepackage{dcolumn}
\usepackage{bm}

\usepackage[caption=false]{subfig}
\usepackage{braket}

\begin{document}

\preprint{}

\title{Finite-temperature real-time properties of magnetic polarons in two-dimensional quantum antiferromagnets}


\author{Toni Guthardt}
\affiliation{\LMUMunich}

\author{Markus Scheb}%
\affiliation{\LMUMunich}

\author{Jan von Delft}
\affiliation{\LMUMunich}
\affiliation{\MCQST}

\author{Fabian Grusdt}
\affiliation{\LMUMunich}
\affiliation{\MCQST}

\author{Annabelle Bohrdt}
\affiliation{\LMUMunich}
\affiliation{\MCQST}
\affiliation{\Regensburg}



\date{\today}

\begin{abstract}
Due to significant progress in quantum gas microscopy in recent years, there is a rapidly growing interest in real-space properties of single mobile dopands created in correlated
antiferromagnetic (AFM) Mott insulators. 
However, a detailed numerical description remains challenging, even for simple toy models. As a consequence, previous numerical simulations for large systems were largely limited to $T=0$. To provide guidance for cold-atom experiments, numerical calculations at finite temperature are required. Here, we numerically study the real-time properties of a single mobile hole in the 2D $t$-$J$ model at finite temperature and draw a comparison to features observed at $T=0$. 
We find that a three-stage process of hole motion, which was reported at $T=0$, is valid even at finite temperature. However, already at low temperatures, the average hole velocity at long times is not simply proportional to the spin coupling, contrary to the $T=0$ behavior.
Comparing our finite-temperature numerical results with the experimental data from  
quantum gas microscopy we find a qualitative disagreement: in experiment, 
hole spreading speeds up with increasing $J/t$, while in 
our numerics it slows down.
The latter is
consistent with the numerical findings previously reported at $T=0$.
\end{abstract}

\maketitle

\section{Introduction}\label{sec:introduction}

The parent compound of cuprate superconductors is believed to be a two-dimensional Heisenberg antiferromagnet (AFM)   \cite{Lee2006highTc}. In addition, it is generally assumed that an interplay between hole motion and antiferromagnetism is at the heart of high-temperature superconductivity in cuprates \cite{proust2019remarkable}. Therefore, it is of great interest to study the behavior of a single mobile charge carrier in an
antiferromagnetic spin background \cite{kanasz2017quantum,nielsen2022nonequilibrium,nielsen2021spatial,vijayan2020time,kurokawa2023unveiling,qiao2025realization,bohrdt2024microscopy,Hahn2022spinchargedeconfinement,Ji2021Apr}, forming a so-called magnetic polaron.

Previous theoretical studies of 
the real-space motion of a hole in an AFM spin background \cite{shen2024finite,bohrdt2020dynamical,Grusdt2018Parton,nielsen2022nonequilibrium}  have shown that the dynamics at $T=0$ follows a three-stage process,
involving ballistic hole spreading, the emergence of a polaron, and ballistic polaron spreading. However, the behavior at finite temperature is not yet fully understood.
In addition to numerical simulation \cite{bohrdt2020dynamical,bohrdt2020parton,elser1990ground,dagotto1990strongly,leung1995dynamical,brunner2000single,mishchenko2001single,white2001density,sangiovanni2006static,mezzacapo2011variational} various semi-analytical and variational approaches \cite{kane1989motion,sachdev1989hole,bermes2024magnetic,schmitt1988spectral,martinez1991spin,auerbach1991small,liu1992dynamical,boninsegni1992green,boninsegni1992variational,starykh1995self} have been employed to investigate polaron formation.
In the conventional magnetic polaron picture, the polaron can be understood as a cloud of correlated magnons dressing
the hole \cite{kane1989motion,sachdev1989hole,shen2024finite,bermes2024magnetic}. It has been shown that a parton picture \cite{Grusdt2018Parton,bohrdt2020parton,manousakis2007string,grusdt2019microscopic}, 
first suggested by B\'eran et al. \cite{beran1996evidence} for describing the underlying magnetic polaron, is able to capture the relevant physics qualitatively. Here, a magnetic polaron consists of a holon, carrying the charge, which is connected by a string of displaced spins with a spinon, carrying the spin \cite{Grusdt2018Parton,bohrdt2020dynamical,beran1996evidence}. In one of the simplest approximate descriptions of a magnetic polaron in the parton picture the so-called frozen spin approximation (FSA) is used, which considers only charge fluctuations along strings of displaced spins and leaves the wavefunction of the surrounding spins unaffected by hole hopping \cite{Grusdt2018Dec,bohrdt2020dynamical,grusdt2019microscopic}. In this work, we aim to test to which extent predictions of this parton model are valid at finite temperature.

In a complementary line of work, there has been
significant recent progress in the study of the real-space properties of magnetic polarons experimentally using quantum gas microscopy (QGM) \cite{bohrdt2021exploration,gross2017quantum}.
This progress has enabled the first real-space observation of magnetic polarons in equilibrium \cite{koepsell2019imaging} and out of equilibrium \cite{Ji2021Apr}. By employing this technique, it is possible to perform large-scale two-dimensional simulations at finite temperature and to study both real-space and time properties. 
These finite-temperature simulations are a first step towards understanding the intriguing finite-temperature phases observed in cuprates. Hence, it is essential to perform numerical calculations at finite temperature in order to provide guidance for QGM.

Here, we report on the numerical simulation of the real-space, finite-temperature dynamics of a hole in a four-leg cylinder described by the $t$-$J$ model, see Fig. \hyperref[fig:figure_greiner_comparison]{1(a),(b)}. First, we observe that the main stages of hole motion reported at $T=0$ \cite{shen2024finite,bohrdt2020dynamical,bohrdt2020parton,Grusdt2018Parton,nielsen2022nonequilibrium} are valid even at finite temperature. However, contrary to the $T=0$ behaviour, the hole velocity observed for times larger than $1/J$ is not proportional to the spin coupling $J$. This can be seen already at temperatures $T=0.5J$ and above. 
Second, a comparison of our numerical results with experimental data from QGM reveals short time agreement, see Fig. \hyperref[fig:figure_greiner_comparison]{1(c)}
(and Sec.~\ref{subsec:experiment_comparison} for a detailed discussion). However, starting at intermediate times, we observe disagreement in the behavior of the $J/t$ dependence of hole dynamics:  in experiment, the hole spreading speeds up with increasing $J/t$, whereas for our numerical results it slows down. The latter 
is in line with the numerical behavior previously reported at $T=0$ \cite{bohrdt2020dynamical,nielsen2022nonequilibrium} and demonstrates the necessity for systematic studies and increased comparison between experiment and numerical simulation.

The structure of this paper is as follows: In \Sec{sec:methods}, we present the numerical details of our tensor network simulation. This is followed in \Sec{sec:results} by an analysis of the results of our real-time dynamics at finite temperature: We start with an introduction to the background knowledge, then illustrate the behavior of the dynamics when varying temperature and coupling ratio. This is followed by a discussion of spin correlations. Finally, we conclude this section on dynamics with a short comparison to experiment. We close in \Sec{sec:SummaryOutlook} by discussing implications of our work and future research directions.

\section{Model and numerics}
\label{sec:methods}

It is generally accepted that the Fermi-Hubbard-model provides a good starting point for a theoretical description of cuprates \cite{emery1987theory,dagotto1994correlated,Lee2006highTc}. At strong coupling, it can be mapped to the $t$-$J$ model up to $\mathcal{O}(t^3/U^2)$,
\begin{equation}\label{eq:tjmodel}
\hat{H} = -t\sum_{\braket{\mathbf{i}\mathbf{j}},\sigma}\mathcal{\hat{P}}(\hat{c}^{\dagger}_{\mathbf{i},\sigma}\hat{c}_{\mathbf{j},\sigma} + \mathrm{h.c.})\mathcal{\hat{P}} + J \sum_{\braket{\mathbf{i}\mathbf{j}}}\left(\mathbf{\hat{S}_i} \cdot \mathbf{\hat{S}_j}-\frac{\hat{n}_\mathbf{i} \hat{n}_\mathbf{j}}{4}\right), 
\end{equation}
where the first term denotes the hopping of holes with amplitude $t$ and the second term represents the spin-exchange interactions with coupling constant $J =4t^{2}/U$. Note that $\mathcal{\hat{P}}$ projects onto the space with at most one fermion per site and we neglected a three site term \cite{spalek2007tj} in \Eq{eq:tjmodel}. 
Despite the apparent simplicity of this model, theoretical predictions and numerical simulations have proven challenging.
As a consequence, previous theoretical calculations of hole dynamics have been limited to $T=0$ behavior or required additional approximations in order to reach finite temperature \cite{shen2024finite}. 

\begin{figure}
\centering
\includegraphics[width = 0.48\textwidth]{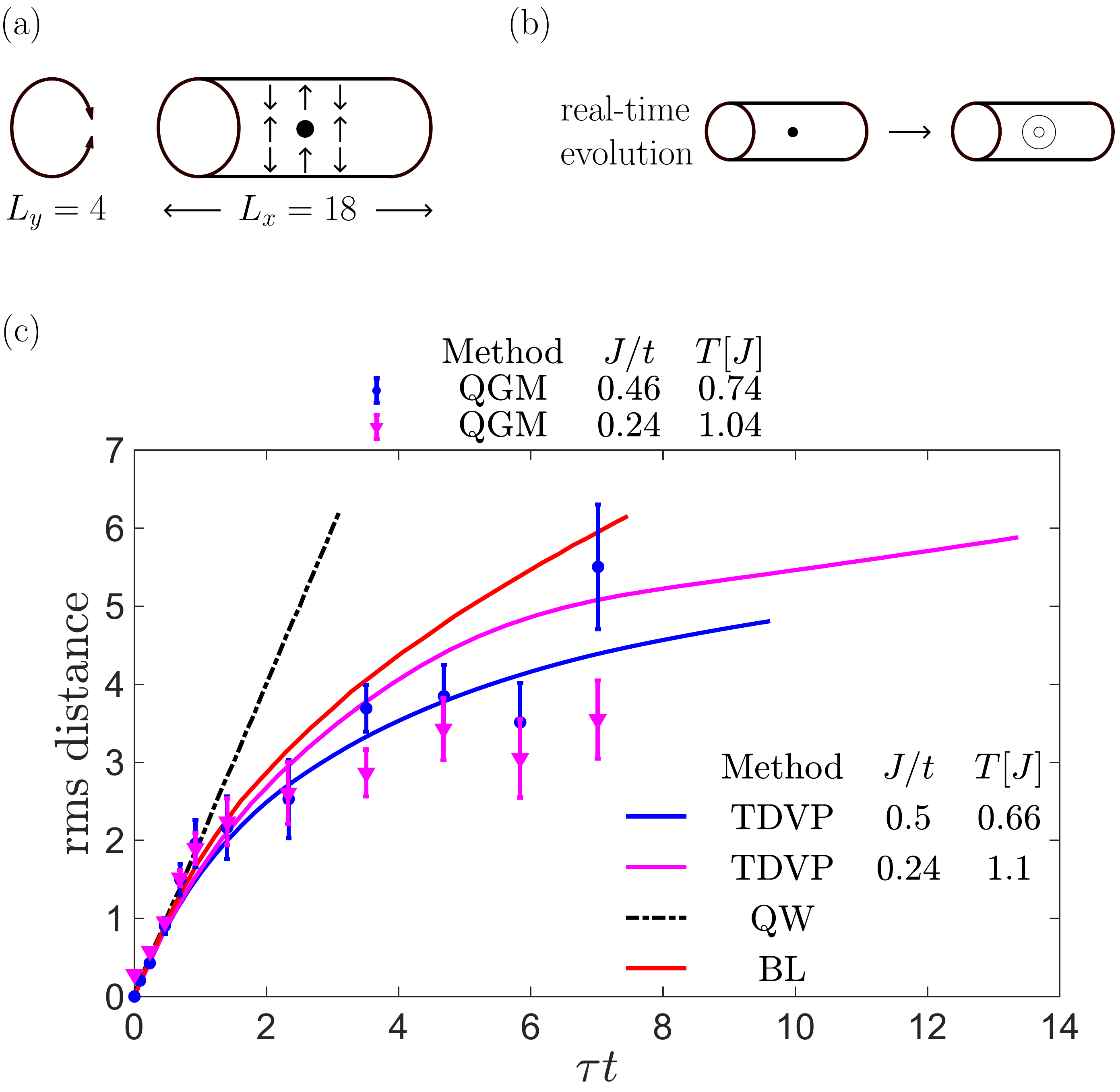}
\caption{(a) Illustration of the cylinder used in the simulation. The system is displayed in the configuration present at the beginning of the time evolution. The black circle represents the hole density and the arrows correspond to spin. (b) Illustration of the real-time spreading of the hole. (c) Comparison of numerical data with experimental  QGM results \cite{Ji2021Apr}. The plot displays the root-mean-square (rms) hole distance (defined in \Eq{eq:d_rms-extrapolated}) as a function of time $\tau$. The numerical data computed on a cylinder was extrapolated to a plain 2D lattice to allow for better comparison (see Sec.~\eqref{subsec:experiment_comparison}). The initial linear increase signifies a ballistic expansion that is consistent with a non-interacting quantum walk (QW) \cite{Ji2021Apr}. The subsequent slowdown of the hole can be approximated by an analytical model which is based on a free quantum walk in a Bethe lattice (BL) \cite{Ji2021Apr}. The data is displayed for two different $J/t$ values.}
\label{fig:figure_greiner_comparison}

\end{figure}

All of our simulations were prepared by calculating the thermal equilibrium of the $t$-$J$-model at half-filling on a cylinder with length $L_x = 18$ and width $L_y=4$, see Fig. \hyperref[fig:figure_greiner_comparison]{1(a)}. To that end, we used the density matrix renormalization group (DMRG) \cite{White1992Nov,White1993Oct} in the language of matrix-product-states (MPSs) \cite{Schollwoeck2011dmrg}, adapted to finite temperatures via a purification scheme \cite{Nocera2016puri,Feiguin2005puri,Feiguin2010puri} and enhanced by the use of disentangling algorithms \cite{Hauschild2019disent}. The resulting system exhibits insulating antiferromagnetic properties. We then modified it by removing a single fermion, thus enabling the subsequent motion of a hole and formation of a magnetic polaron, see Fig. \hyperref[fig:figure_greiner_comparison]{1(b)}.
The subsequent dynamics were simulated by combining two versions of MPS-based time evolution algorithms \cite{paeckel2019timeevol}. Although the entanglement physically only spreads locally around the location of the quench, it is encoded via the virtual bonds of the MPS, which spans the entire lattice. This is why we began the time evolution with a single step of the more expensive, but global Krylov scheme \cite{Garcia-Ripoll2006Dec,Dargel2012May,Wall2015}. The rest of the time evolution was performed via the local, but less expensive time-dependent-variational-principle (TDVP) algorithm \cite{Haegeman2011TDVP,Haegeman2016TDVP}. This procedure was improved by the use of a backwards-time-evolution scheme \cite{Karrasch2012anc,Kennes2016anc,Karrasch2013anc}, which allowed us to reach longer times without additional approximations. In all of the above algorithms, we also used controlled bond expansion \cite{Gleis2023CBE,Li2024TDVP}, which effectively performs two-site optimizations at one-site costs. To guarantee the accurate implementation of the code, we conducted benchmark tests on the exactly solvable $xy$ chain and the non-interacting tight binding chain/cylinder.

\section{Real time dynamics}
\label{sec:results}

In this section we shed light on the real-time dynamics of a single hole at different temperatures and coupling ratios. In the following we also draw a comparison with the behaviour observed at $T=0$ and experimental results from QGM. We denote the time variable by $\tau$, and plot it in units of either inverse tunneling $1/t$ or inverse exchange coupling $1/J$.

\label{subsec:dynamics}

\begin{figure}
\centering
\includegraphics[width = 0.48\textwidth]{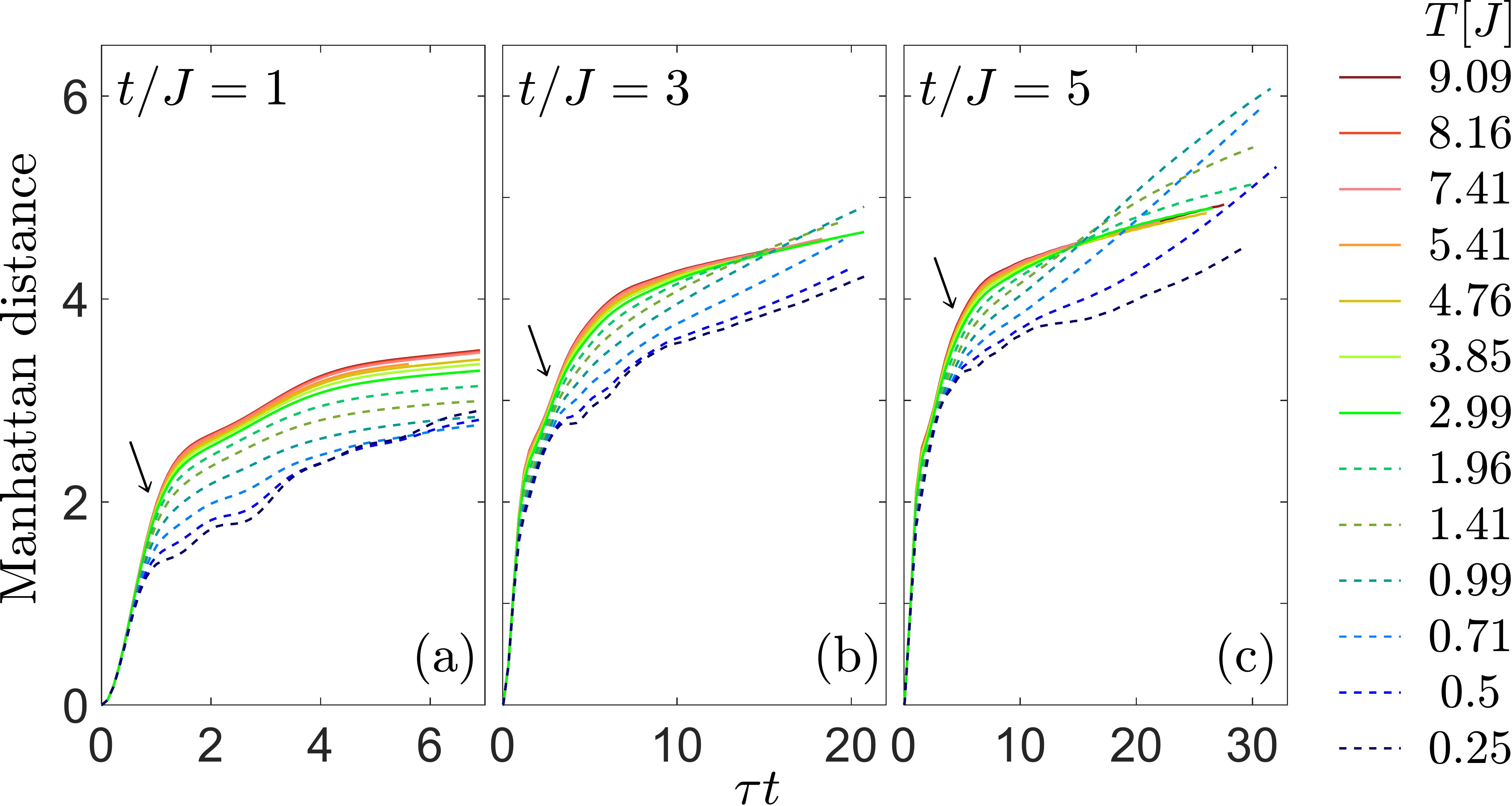}
\caption{Dynamics of a hole at different temperatures (colored lines) in the two-dimensional $t$-$J$ model on a square lattice. The calculations are performed for $t/J =1,3,5$ on a cylinder with length $L_x = 18$ and width $L_y=4$. The plots show the Manhattan distance as a function of time $\tau$. For all values of $t/J$ a time of $\tau J \simeq 7$ was reached. At strong coupling, i.e. $t/J \gg 1$, and times larger than $1/J$ we observe faster spreading at low temperatures (dashed lines) than at higher temperatures (solid lines). The arrows indicate when $\tau J =1$.}
\label{fig:figure1}

\end{figure}

\subsection{Background}
\label{subsec:background}
 
Previous theoretical studies at $T=0$ \cite{shen2024finite,bohrdt2020dynamical,bohrdt2020parton,Grusdt2018Parton,nielsen2022nonequilibrium} have shown that the dynamics of a hole follows a three-stage process: (i) Initially, the hole spreads ballistically with a velocity proportional to $t$, independent of $J$, up to time $1/t$. (ii) The magnetic polaron emerges as a meson, consisting of a holon and a spinon. This process can involve damped oscillations, reflecting the structure of the meson. (iii) Starting at times $1/J$, the polaron spreads ballistically with a velocity proportional to $J$ and independent of $t$.

In order to demonstrate the individual stages of the three-stage process, one can analyze the Manhattan distance $r$ \cite{bohrdt2020dynamical}
\begin{equation}
r = \sum_{x} \sum_{y}(|x| + |y|) \cdot n^{h}(x,y) \, ,
\end{equation}
where $x$ and $y$ denote positions within the lattice, and $n^h(x,y)$ the corresponding hole density. The origin, with $(x,y)=(0,0)$, is defined as the initial hole location. By examining the time-dependence of $r$ for $T>0$, we study how far the hole is moving from its original position. Furthermore, this allows us to gain insight into the extent to which the polaron retains its characteristics at higher temperatures.

\subsection{Varying temperature}
\label{subsec:temp_dependent_dynamics}
In \Fig{fig:figure1}, we compare the time evolution of the Manhattan distance $r(\tau)$ for various temperatures $T$, while
keeping the coupling ratio $t/J$ fixed. For all values of $T$ and $t/J$, we
observe the expected behaviour of an initial fast spreading of the hole, followed by a slower propagation due to magnetic dressing. Note that this corresponds to a three-stage process, similar to the three stages (i-iii) reported at $T=0$, see \Sec{subsec:background}. For an analysis of the extent to which the stages (i-iii) are still present at finite temperature, see \Sec{subsec:three_stage_process}.

In the following, the term monotonic $T$-dependence is referred to as an increase/decrease of $r(\tau)$ at a fixed $\tau$ when the temperature increases/decreases.
Upon closer examination, we notice that the onset of non-monotonic $T$-dependence occurs at longer times $\tau J$ as $t/J$ is decreased.
For stronger coupling, $t/J>1$, low-temperature values of $r$ (dashed lines) stay lower for times up to $1/J$ (indicated by arrows), i.e. up to the stage of polaron formation, but start to increase more quickly at times, $\tau \approx 3/J$. As a result, $r(\tau)$ at fixed $\tau > 1/J$ is larger at intermediate temperatures than at large ones. This effect is more pronounced the higher $t/J$.

In the parton picture, spin-spin-correlations are necessary for a finite string tension that constrains hole expansion and binds the holon to the spinon. At short times up to $\tau \approx 1/J$ the reduction in string tension with $T$ results in a monotonic dependence of $r(\tau)$ on $T$. This is due to the fact that the spinon motion can be neglected during this period. This monotonic $T$-dependence can be observed in \Fig{fig:figure1}. However, at longer times, the spinon starts to move ballistically at low temperatures. Hence, at low $T$, the entire magnetic polaron propagates faster. This also provides an explanation for the non-monotonic dependence of $r(\tau)$ on $T$ reported above.

We will further elaborate this point in the next subsection.

\begin{figure}
\centering
\subfloat{
\includegraphics[width = 0.47\textwidth]{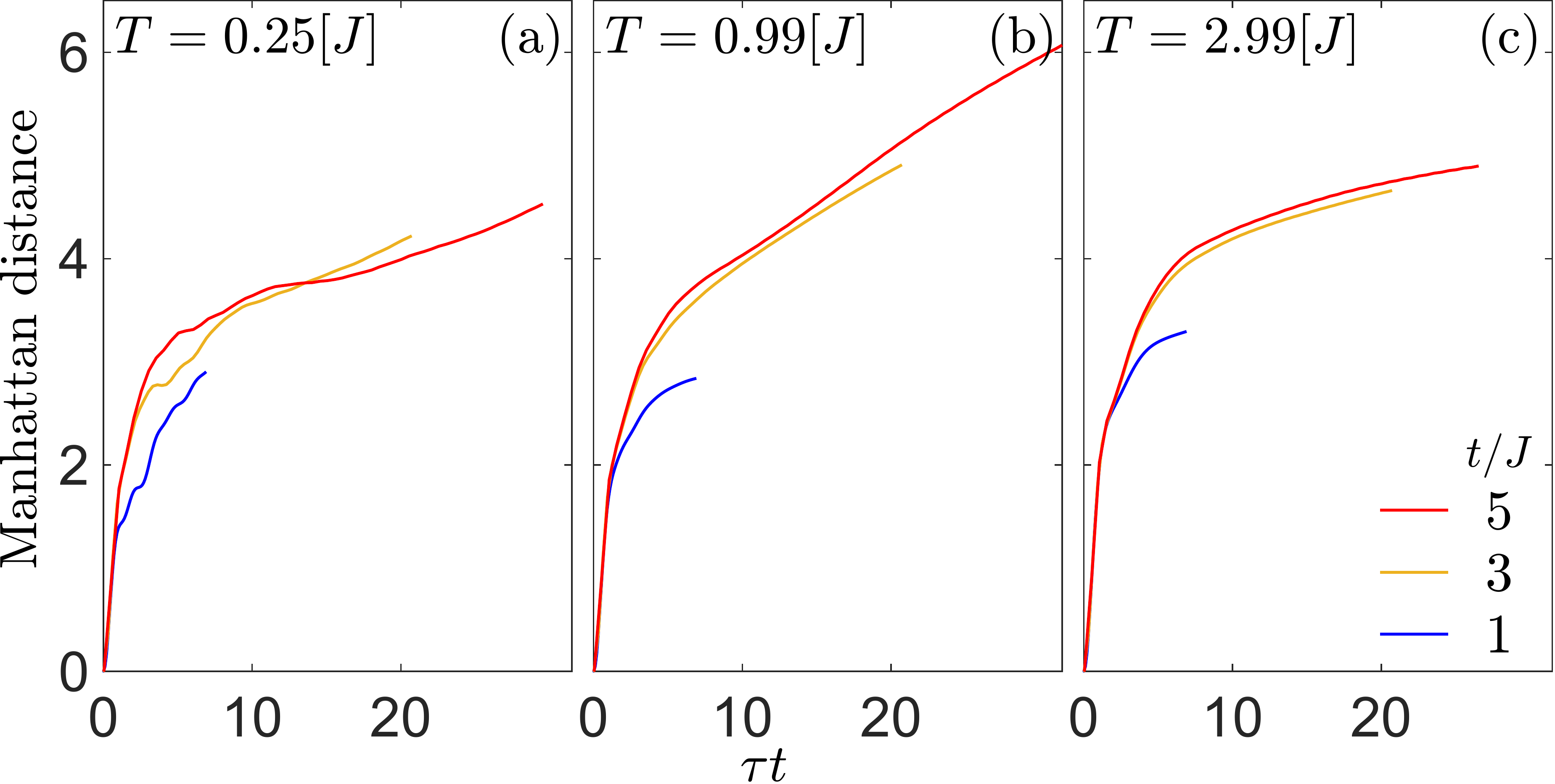}\label{fig:figure2a-c}}

\vspace{-0.15cm}

\subfloat{
\includegraphics[width = 0.47\textwidth]{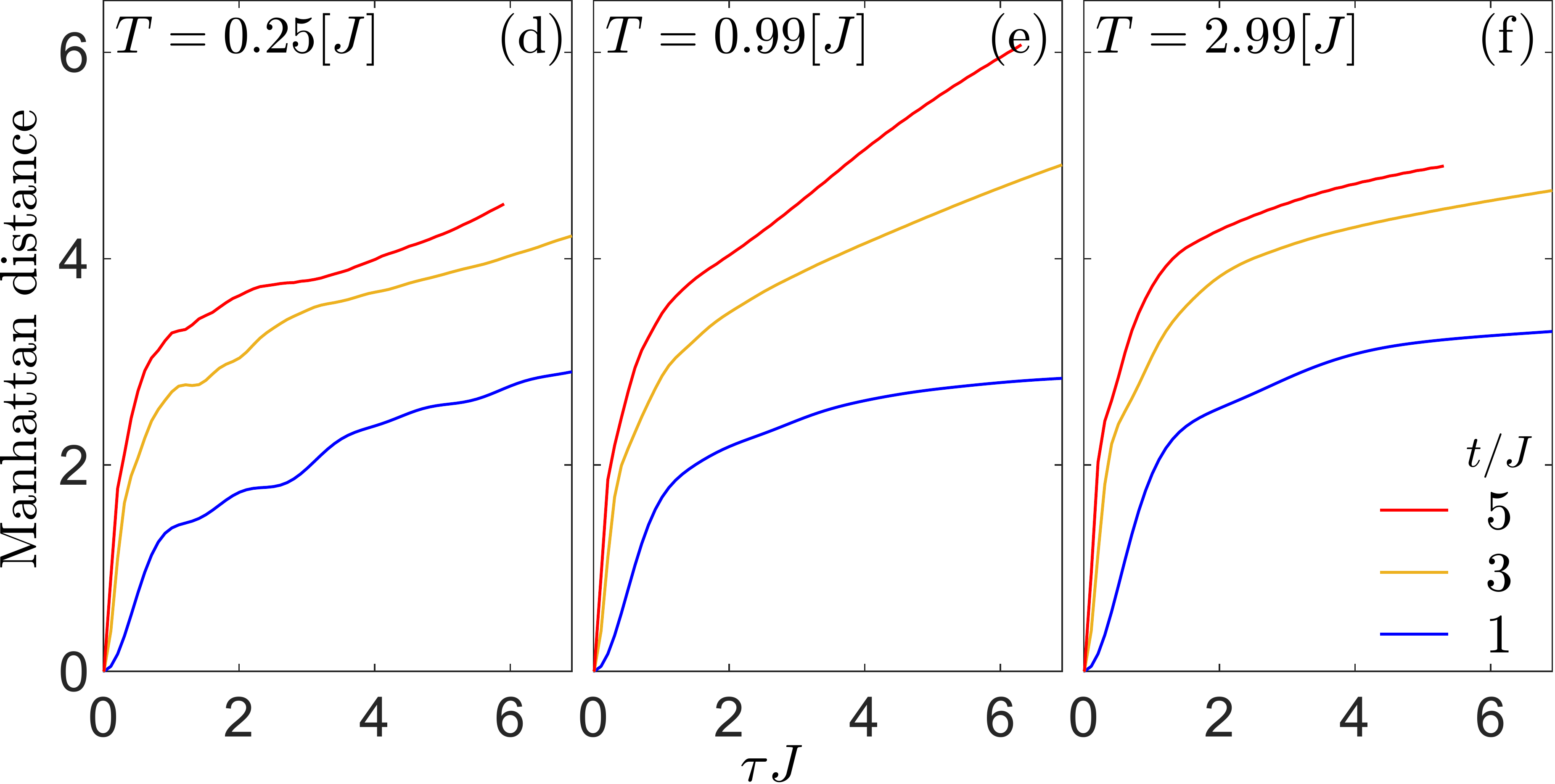}\label{fig:figure2d-f}}

\caption{Analysis of the three-stage process at finite temperature. (a-c) Manhattan distance is shown for three values of $t/J$ at several temperatures as a function of time $\tau$, plotted in units of $1/t$. (d-f) Analog to (a-c), but now plotted in units of $1/J$. The initial spreading still occurs with a velocity proportional to hopping $t$. For times beyond the emergence of the polaron we do only observe a simple proportionality of the expansion rate to the spin coupling $J$ at low temperature $T=0.25J$.}
\label{fig:figure2}

\end{figure}

\subsection{Varying the coupling ratio}
\label{subsec:three_stage_process}
In order to analyze the velocity of hole propagation, we compare the time evolution of the Manhattan distance $r(\tau)$ for three values of $t/J$ while keeping the temperature fixed, see \Fig{fig:figure2}. The curves in Fig. \hyperref[fig:figure2a-c]{3(a-c)} and Fig. \hyperref[fig:figure2d-f]{3(d-f)} represent the same data, but for different scalings of the time axis, $\tau t$ or $\tau J$, focusing on the short- or long-time dynamics of the hole, respectively.

Fig. \hyperref[fig:figure2a-c]{3(a-c)} reveals that the velocity initially does not depend on the spin-coupling $J$ but only depends on hopping $t$, as can be seen by all curves lying on top of each other for shorter times. For all $t/J$ values, an increase of temperature from $T=0.25J$ to $T=2.99J$ leads to an extension of the first stage of hole propagation. We interpret this as arising from a reduction in spin correlations, resulting in a reduction in string tension, with increasing temperature. Consequently, the time window within which the hole expansion only depends on hopping $t$ increases. This effect is most pronounced for $t/J=1$.

By taking a closer look at Fig. \hyperref[fig:figure2d-f]{3(d)}, featuring $T=0.25J$, we notice that the three curves are approximately 
linear and parallel for larger times. Since the velocity can be estimated by dividing the Manhattan distance by time, and time is scaled in units of $1/J$, all three curves having the same slope verifies that the polaron velocity is proportional to $J$. However, upon increasing the temperature above $T=0.25 J$, the long time behaviour starts to differ. For $T=0.99 J$, see Fig. \hyperref[fig:figure2d-f]{3(e)}, the different graphs, no longer run parallel to each other at large times, implying a deviation from the behavior found for the polaron model at $T=0$.
For $T=2.99J$, see \hyperref[fig:figure2d-f]{3(f)}, the non-parallel behaviour is less pronounced, but the three graphs are significantly more curved than for $T=0.25J$. 

Furthermore, the findings from Fig. \hyperref[fig:figure2d-f]{3(d-f)} are reinforced from a different perspective. By scaling time in units of $1/t$, see Fig. \hyperref[fig:figure2a-c]{3(a-c)}, and increasing $t/J$, we effectively decrease $J$ and expect a decrease of the velocity for longer times, at least for lower temperatures. This is visible in the $T=0.25 J$-plot by the crossing of the curves for $t/J=5$ (red) and $t/J=3$ (yellow). For $T=0.99J$ and $T=2.99J$, this behaviour is absent, pointing again to a different behaviour of polarons at longer times.

\begin{figure}
\centering
\subfloat{
\includegraphics[width = 0.48\textwidth]{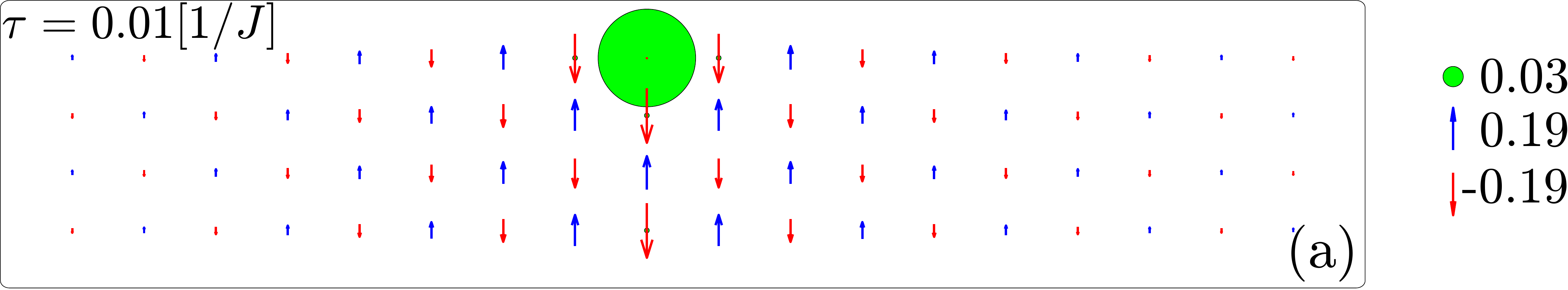}\label{fig:figure4a}}

\vspace{-0.15cm}

\subfloat{
\includegraphics[width = 0.48\textwidth]{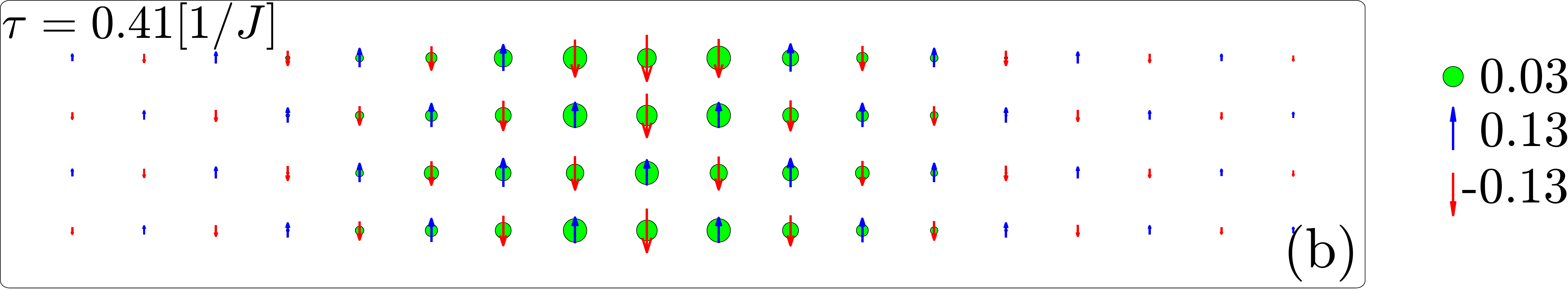}\label{fig:figure4b}}

\vspace{-0.15cm}

\subfloat{
\includegraphics[width = 0.48\textwidth]{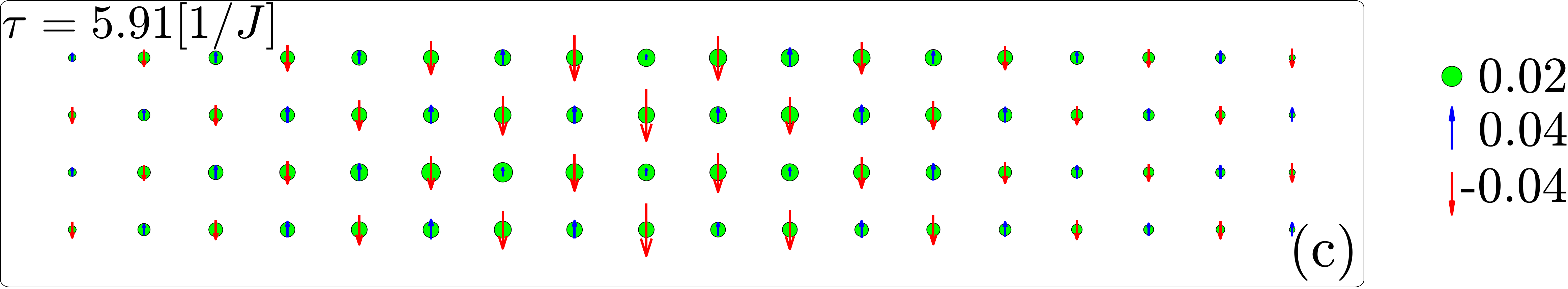}\label{fig:figure4c}}

\caption{Hole density and spin across the entire lattice shown for several times at temperature $T=0.25J$ for $t/J=5$ . The size of the green circles represents the value of the hole density. Height and direction of the arrow correspond to absolute value and direction of spin.}
\label{fig:figure4}

\end{figure}

\begin{figure}
\centering
\subfloat{
\includegraphics[width = 0.48\textwidth]{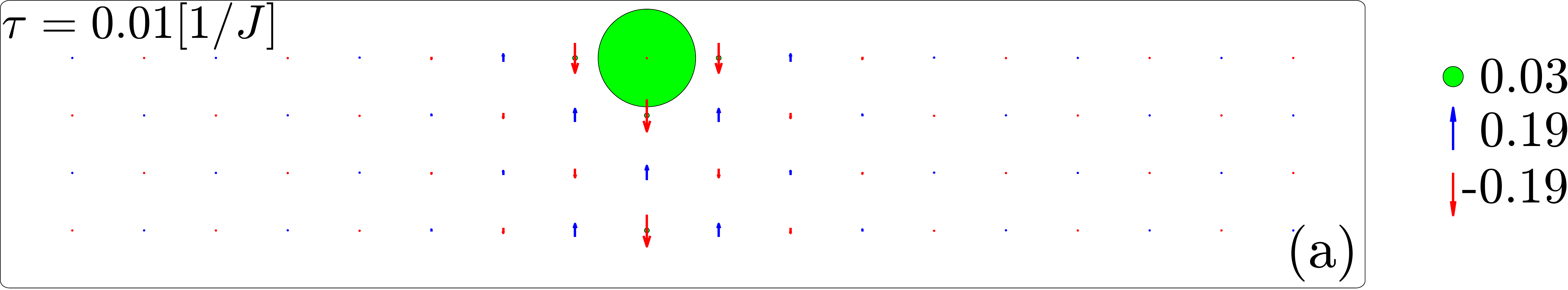}\label{fig:figure5a}}

\vspace{-0.15cm}

\subfloat{
\includegraphics[width = 0.48\textwidth]{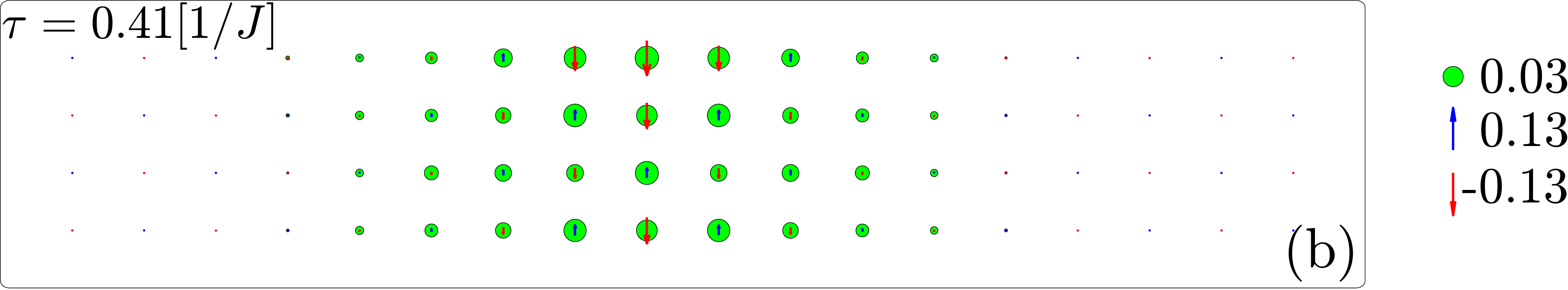}\label{fig:figure5b}}

\vspace{-0.15cm}

\subfloat{
\includegraphics[width = 0.48\textwidth]{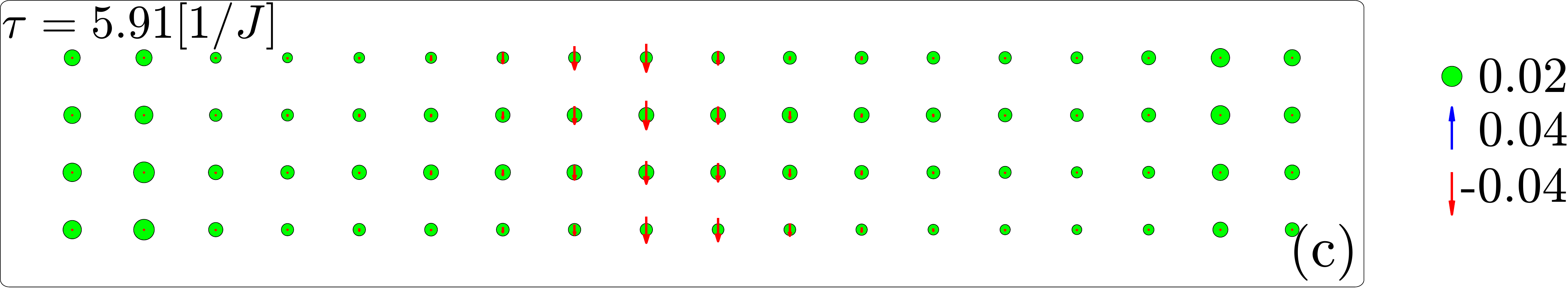}\label{fig:figure5c}}

\caption{Same as \Fig{fig:figure4}, but for $T=0.99J$ and $t/J=5$.}
\label{fig:figure5}

\end{figure}

\subsection{Hole density and spin across system}
\label{subsec:hole_dens_and_spin}

Thus far, our analysis has been limited to average distances. We now turn to site-resolved densities, which are directly accessible in quantum gas microscopes.
In the following, we illustrate how both the hole density and spin evolve as a function of the lattice sites and as a function of time for a specific temperature $T=0.25J$ with $t/J=5$, see \Fig{fig:figure4}. At this point it is also important to mention that initially, an electron with spin down was removed from the equilibrium system, resulting in a total spin $S_z^{\mathrm{tot}} \neq 0$.

In \Fig{fig:figure4a} we observe how the short-time symmetric spreading of the hole, starting at the initial hole position, results in spins being aligned in the same direction at sites adjacent to the initial hole position. This observation reflects that $S_z^{\mathrm{tot}} \neq 0$. The initial hole position is located in the center of the cylinder and corresponds to the site with the largest hole density present at such short times. This indicates that the hole is still mainly located at the initial site. 

At intermediate times, see \Fig{fig:figure4b}, the hole has already spread over one third of the length of the cylinder. In the process of spreading it has distorted the spin order around the initial hole position significantly more compared to \Fig{fig:figure4a}. It is important to point out that the parallel alignment of spins found at the initial hole location is characteristic of a spinon. 

Displacing spins in an AFM background comes at an energy cost. This slows the initial fast spreading of the hole at times $1/t$, which we observe here.

Finally at long times, see \Fig{fig:figure4c}, we can see the reemergence of AFM correlations in the whole system and an almost uniform distribution of the hole density over the entire lattice, indicating the return of the system into equilibrium.

In combination, \Fig{fig:figure4} is a direct demonstration of the three-stage process in real-space, (i) starting with the fast hole spreading in \Fig{fig:figure4a}, (ii) followed by the magnetic polaron formation in \Fig{fig:figure4b} and (iii) concluding with the spinon spreading in \Fig{fig:figure4c}.

\Fig{fig:figure5} presents analogous results at a higher temperature $T=0.99[J]$. One observes a similar behaviour compared to \Fig{fig:figure4}, but the hole motion generally takes place faster and the average spin expectation value is reduced significantly. Nevertheless, the spatial heterogeneity of the hole density at long times indicates that the system has not yet reached a state of near-equilibrium, see \Fig{fig:figure5c}. This phenomenon may be attributed to finite size effects. At the same point in time one also observes a slight asymmetry in the spin data with respect to mirroring the data along the initial hole position. This presumably is due to accumulation of errors in the numerical simulation, which affect the rather small spins at such temperatures more severely. Furthermore, we can see that no AFM spin patterns have built up at our longest time.

\begin{figure}
\centering
\includegraphics[width = 0.48\textwidth]{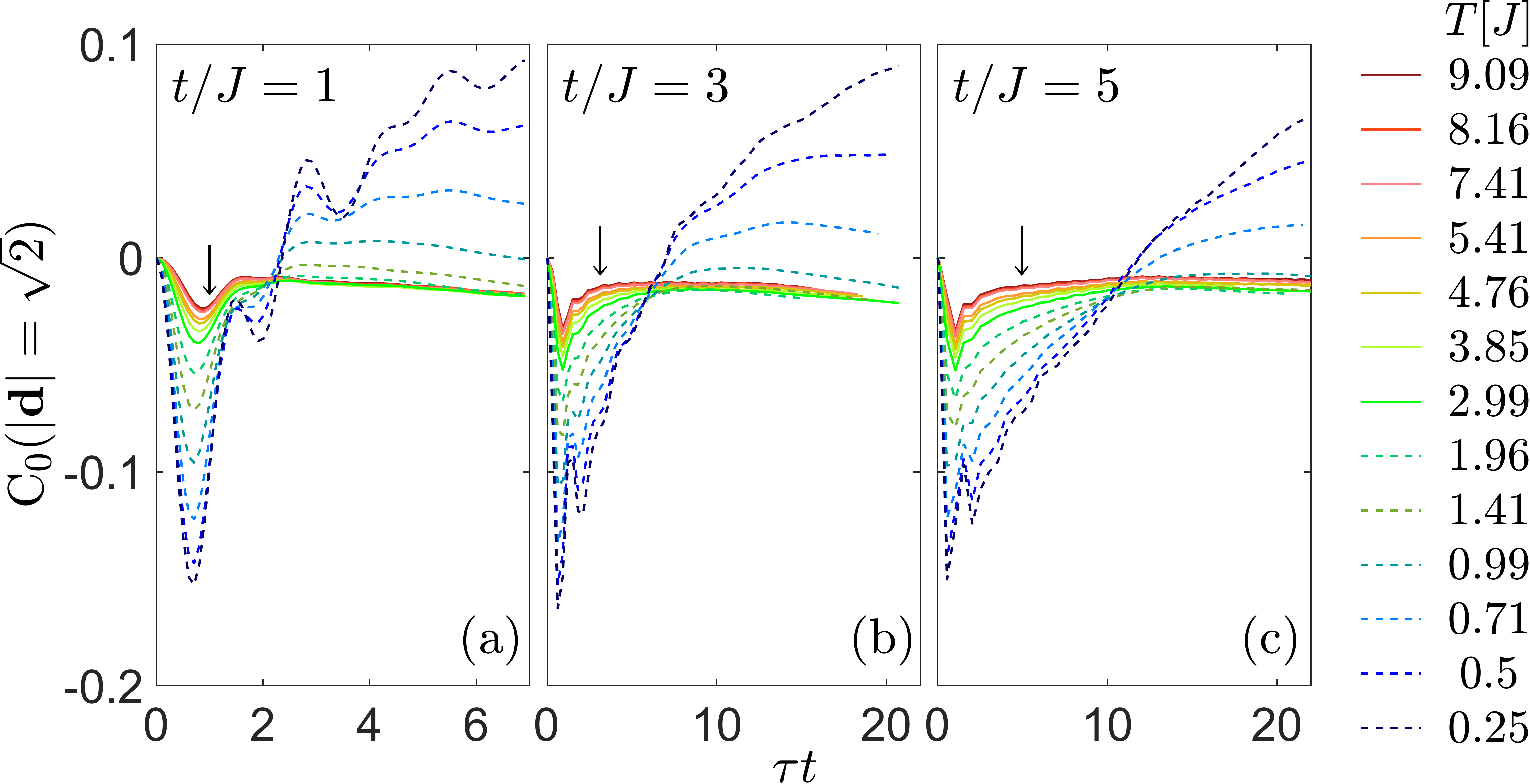}
\caption{Staggered next nearest-neighbor spin correlations between the initial hole position and its diagonal neighbors as function of time. Data is displayed for several $t/J$ at different temperatures. The arrows indicate when $\tau J =1$.}
\label{fig:figure6}

\end{figure}

\subsection{Next-nearest-neighbor spin correlations}
\label{subsec:next_nn_spin_correl}

The process of polaron creation and subsequent polaron spreading can also be analyzed by considering the evolution of spin correlations. In equilibrium, i.e. directly before we remove an electron from our system and study its dynamics, we can observe non-negligible local AFM correlations for temperatures $T \lesssim 1J$, see \App{appendix:spin_correl_in_equil}.

To gain further insights into the evolution of spin correlations, we study staggered spin correlations, which are evaluated between sites $\mathbf{r}$ and $\mathbf{r'}$ and defined as
\begin{equation}\label{eq_sign_corrected_spin_correl}
C_{\mathbf{r}}(\mathbf{d})=(-1)^{d_x + d_y}4(\braket{\hat{S}^{z}_{\mathbf{r}} \hat{S}^{z}_{\mathbf{r} + \mathbf{d}}} - \braket{\hat{S}^{z}_{\mathbf{r}}} \braket{\hat{S}^{z}_{\mathbf{r} + \mathbf{d}}} ).
\end{equation}
Here, $\mathbf{d}$ is defined as the difference vector between respective sites $\mathbf{r}$ and $\mathbf{r'}$ and $\hat{S}^{z}$ is defined as the usual $z$-component of the spin operator $\mathbf{\hat{S}}$. Note that as a consequence of this definition, positive correlation values correspond to AFM correlations. 

We start by considering the corresponding next-nearest-neighbor correlations $C_0(|\mathbf{d}|=\sqrt{2})$. In \Fig{fig:figure6} we display the staggered next-nearest-neighbor spin correlation (SNNNC) as a function of time and relative to the initial hole position. This has been a common choice in experiment \cite{Ji2021Apr}, facilitating comparisons. The correlations are evaluated for different temperatures at three values of $t/J$. 

We observe for all values of $t/J$ at all temperatures that the system is out of equilibrium during short and intermediate times, testified by the presence of negative correlation values, and only slowly approaches a steady state for long times. For times up to $1/t$, $C_0(|\bm{d}|=\sqrt{2})$ is negative and increases even further in magnitude. This is connected to the fast initial spreading of the hole with a velocity proportional to $t$. When the hole performs one hop, it places the neighboring spin on the "origin". As a result, the spin is situated in the "wrong" sublattice, leading to negative $d = \sqrt{2}$ spin correlations. During the phase of polaron emergence, which occurs for times $1/t < \tau < 1/J$, the SNNNC approaches zero quickly and only slows down when the polaron is fully formed. At times larger than $1/J$, when the polaron is moving as a whole, the SNNNC continues to relax to equilibrium, but slower compared to the stage of polaron emergence. Given that the polaron is moving as a whole, the relaxation to equilibrium can also be understood as a consequence of spinon motion away from the origin.

Furthermore we can see that at $T < 2J$ (dashed lines) the negative correlations relax to zero 
much faster for strong spin coupling, i.e. small $t/J$. This phenomenon can be attributed to the relaxation of spin correlation at the origin, which is a consequence of the motion of the spinon away from the origin. Given that spinon motion occurs on time scales of $1/J$, the findings presented on time scales of $1/t$, see \Fig{fig:figure6}, can be explained. 

Since in equilibrium a finite string length of the polaron requires finite spin correlations \cite{bohrdt2020dynamical,grusdt2019microscopic,Grusdt2018Parton}, it is also of interest to determine the temperature up to which finite spin correlations are visible. This should indicate the transition to a region where no polarons emerge. We find an absence of spin correlations close to the origin at all times for temperatures above $T \approx 2J$, with non-vanishing positive spin correlations only present at long times for temperatures smaller $T \approx 1 J$. This temperature scale is in line with the temperature at which one would expect to see a return of AFM correlations due to a competition between temperature and spin coupling. In addition, the complete lack of spin correlations for temperatures above $T=2J$ also suggests that magnetic polarons do not survive in that temperature range.

\begin{figure}
\centering
\includegraphics[width = 0.48\textwidth]{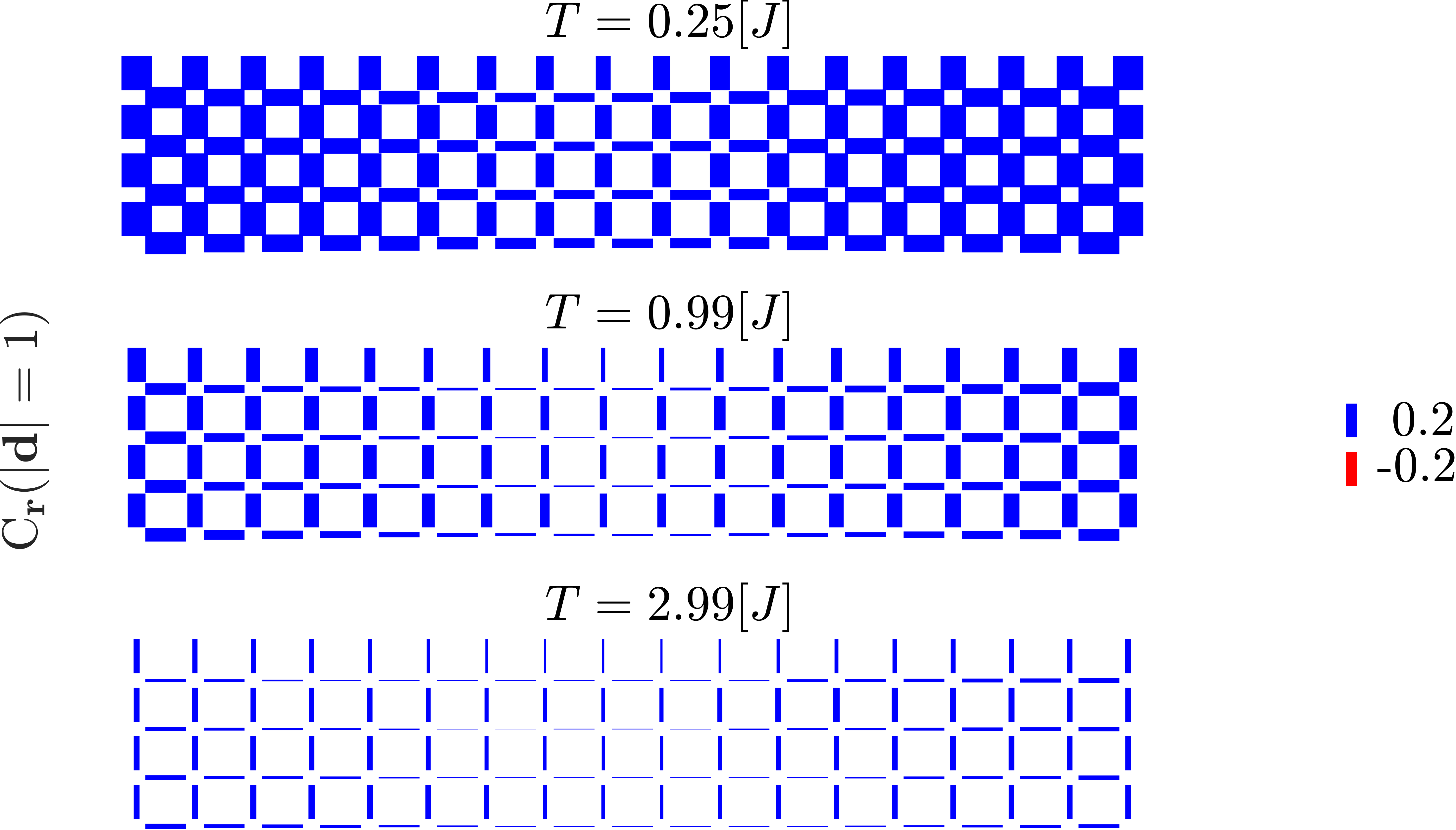}
\caption{Staggered nearest-neighbor (SNNC) spin correlations in the entire lattice for $t/J=5$. Data is plotted at different temperatures at the maximum calculated time $\tau = 6/J$. The bars represent the SNNC spin correlations connecting neighboring sites. The thickness and the color of the bars represents the absolute value and the sign of the spin correlation.}
\label{fig:figure7}

\end{figure}

\subsection{Nearest-neighbor-spin correlations}
\label{subsec:nn_spin_correl}

To further improve our intuitive understanding, we now build on the scenario described in \Sec{subsec:hole_dens_and_spin} by discussing the distribution of spin correlations over the entire lattice as a function of temperature. 

To this end, we take a closer look at the staggered nearest-neighbor (SNNC) spin correlations, see \Fig{fig:figure7}. Here, we present the distribution of the spin correlations over the entire lattice for $t/J=5$ at the maximal reached time $\tau = 6/J$. By examining the long-term correlations, we aim to determine whether we can identify features that are characteristic of a system close to equilibrium, such as a homogeneous spin correlation, or alternatively, whether we can observe features that must be explained due to the dynamics of the doped system.  

Overall, we observe that the average strength of the spin correlation reduces with increasing temperature, as expected. Since these data show that spin correlations have become very weak for temperatures above $T=0.99J$, they support the observations of \Sec{subsec:next_nn_spin_correl}. 

Furthermore, we see a relatively uniform distribution of spin correlation, only the spin correlations around the initial hole position have not yet reached equilibrium. This leads to the conclusion that the system is approaching equilibrium at the end of the time evolution, in agreement with \Sec{subsec:hole_dens_and_spin}.

\subsection{Spinon spreading}
\label{subsec:spinon_spreading}

\begin{figure}
\centering
\includegraphics[width = 0.48\textwidth]{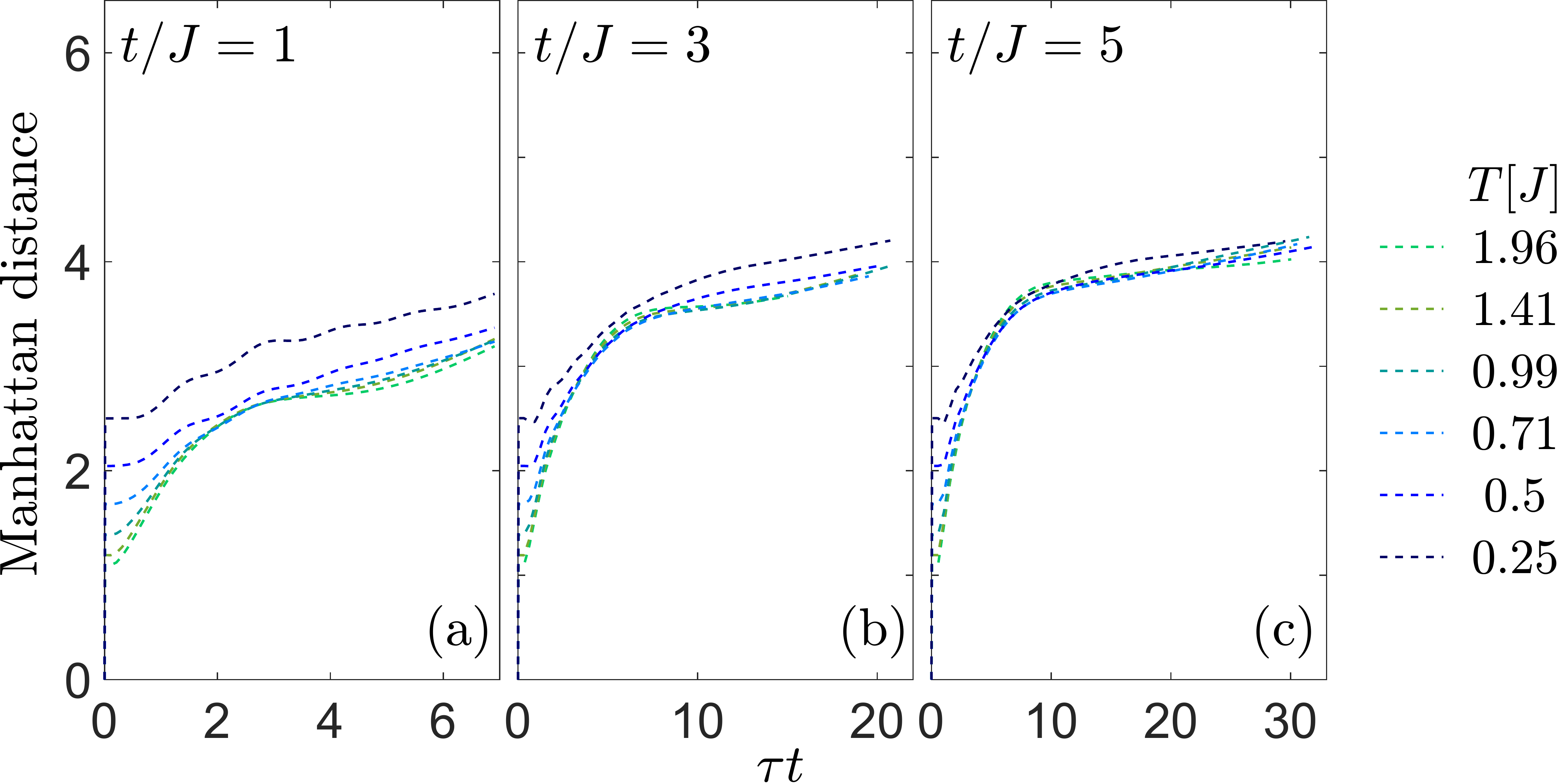}
\caption{Manhattan distance of the spinon. The spinon density used here is defined as the normalized deviation of the SNNC from its equilibrium value. The data is displayed for several $t/J$ and different temperatures, up to which polarons are expected to exist.}
\label{fig:figure_spinon_spreading}

\end{figure}

To conclude the discussion of the temperature dependence, we shed light on the spinon spreading. In order to achieve this, we define the spinon density $n^{s}(x,y)$ as the normalized deviation of the SNNC from its equilibrium value,
\begin{equation}
n^{s}(x,y) = \sum_{|\bm{d}|=1}|C_{(x,y)}(\bm{d}) - C_{(x,y)}^{\mathrm{equil}}(\bm{d})|/h \, ,
\end{equation} 
with $h=\sum_{x,y}\sum_{|\bm{d}|=1}(|C_{(x,y)}(\bm{d}) - C_{(x,y)}^{\mathrm{equil}}(\bm{d})|)$ being the normalization factor. It is crucial to acknowledge that this definition is only valid in the context of temperatures where polarons exist.
\Fig{fig:figure_spinon_spreading} illustrates the Manhattan distance $r_s$ of the spinon for varying values of $t/J$ and temperatures. In analogy to the Manhattan distance $r$ defined for the hole, the Manhattan distance $r_s$ of the spinon is defined as

\begin{equation}
r_s = \sum_{x} \sum_{y}(|x| + |y|) \cdot n^{s}(x,y) \, .
\end{equation}
Overall,  the data indicate that the spinon spreading also follows a three-stage process, similar to the spreading of the hole. 

Within a single short time step $\delta \tau  = 0.01/J$, a pronounced increase in the spinon distance, $r_s$, is observed, with the greatest increase occurring at low temperatures. This can be attributed to the presence of AFM correlations in equilibrium at low temperatures. The removal of an electron and subsequent hopping of the hole for a short time step results in the emergence of ferromagnetic correlations adjacent to the initial hole position. This corresponds to a significant deviation from the equilibrium correlations, which is reflected in a steep increase in the spinon distance $r_s$. 

Following the initial time step, the velocity of spinon spreading is observed to increase with temperature. This behaviour persists up to medium times and is especially pronounced in the case of $t/J>1$. It is noteworthy that this behaviour is consistent with the behaviour of the hole, which also features an increase in spreading velocity with temperature in the case of $T< 2J$ and similar times, see \Sec{subsec:temp_dependent_dynamics}. 

At long times, $\tau > 1/J$, the velocity of spinon spreading is observed to be similar for different temperatures. However, the absolute spinon distance $r_s$ reached at long times is slightly higher for lower temperatures. This is in contrast to the long-time spreading of the hole, see \Sec{subsec:temp_dependent_dynamics}, which reaches further distances the higher the temperature in the case of $T< 1J$ at strong coupling. It is likely that this discrepancy can be attributed to a reduction in the binding strength between the holon and the spinon as the temperature increases, which in turn permits a further propagation of the hole.

\subsection{Comparison with experiment}
\label{subsec:experiment_comparison}

To conclude with the discussion of the hole dynamics we draw a comparison with experimental results using QGM \cite{Ji2021Apr}. To this end, we calculate the root-mean-square (rms) hole distance

\begin{equation}
\label{eq:d_rms}
d_\mathrm{rms} = \sqrt{\sum_{x} \sum_{y}(|x|^2 + |y|^2) \cdot n^{h}(x,y) }\, ,
\end{equation}
where $x$ and $y$ denote positions within the lattice, and
$n^{h}(x,y)$ the corresponding hole density. Analogous to the definition of the Manhattan distance, the origin is defined as the initial hole location. Note that we used $\sum_{x} \sum_{y} n^{h}(x,y) =1$.

The results are presented in Fig.~\hyperref[fig:figure_greiner_comparison]{1(c)} of the introduction. There, we compare the rms distance obtained using our numerics on a cylinder with results from QGM for a plain 2D lattice. Note that the experimental QGM data has been generated for the Hubbard model, whereas our numerical data was computed for the $t$-$J$ model. For better comparison, our numerical results computed on a cylinder were extrapolated to a plain 2D lattice by assuming equivalent long-time expansion in $x$ and $y$ directions
\begin{equation}
\label{eq:d_rms-extrapolated}
d_\mathrm{rms} = \sqrt{\langle x^2 \rangle + \langle y^2 \rangle} \approx \sqrt{2} \sqrt{\langle x^2 \rangle},
\end{equation}
with $\langle x \rangle=\sum_{x} x\cdot n^h(x)$.
The results are displayed for two different $J/t$ values at similar temperatures. For short times up to $\tau \approx 1/t$ we observe good agreement between experiment and numerical simulation. However, starting at intermediate times we find that the $J/t$-dependence observed numerically is contrary to the $J/t$-dependence obtained in experiment: In experiment, the hole spreading speeds up with increasing $J/t$ (blue lines lie lower than red lines), whereas for the numerical results it slows down (blue symbols lie higher than red symbols). Since the numerically observed behavior is in line with previous numerical study at $T=0$ \cite{bohrdt2020dynamical,nielsen2022nonequilibrium}, further analysis is needed to reconcile the experimental measurements with numerical results.

\section{Summary and Outlook}
\label{sec:SummaryOutlook}

In this work, we numerically studied the real-time dynamics of a single hole in the 2D $t$-$J$ model at finite temperature.

We observed that a three-stage process of hole motion previously observed for $T=0$ is valid even at finite temperature. In the strong coupling limit, i.e. $t/J \gg 1$, we observe that the speed of hole spreading decreases with temperature at long times. This suggests strong scattering on thermal excitations, which is not included in the parton model that we use to explain the qualitative behavior at low temperatures. Furthermore, our data shows that the long time spreading is not solely dependent on the spin coupling $J$, indicating that spinons and chargons are no longer bound at high temperatures. 

Furthermore, our findings reveal that, at finite temperature, the initial stage of hole motion is solely dependent on the hopping $t$. Moreover, for all values of $t/J$, an increase in $T$ results in the prolongation of the initial stage of hole motion. 

Additionally, the comparison of our numerical results with experimental data from QGM reveals that the hole spreading speeds up with increasing $J/t$ starting at intermediate times, whereas for our numerical results it slows down. In view of the fact that our findings are in accordance with those of previous numerical analysis performed at $T=0$ in the $t$-$J$ model, we attribute this discrepancy to shortcomings in the theoretical modeling of the experimental setup. It will consequently be intriguing to see whether subsequent studies with QGM can resolve this discrepancy.

Another highly interesting direction for future research would be the investigation of finite doping, e.g. the case of two holes. At $T=0$, both a highly mobile bound pair with a dispersion proportional to $t$ and a heavy pair, which moves due to spin exchange processes, have been found \cite{bohrdt2023twohole}. It will be exciting to see whether similar features can be confirmed at finite temperature and to enhance our understanding of the pairing mechanism in cuprates.

Since our numerical data can be compared directly to QGM data, our research offers guidance for ultracold atom experiments. These experiments have the capability of studying individual magnetic polarons at finite temperature in both real and frequency space. It will therefore be fascinating to see whether a systematic experimental study of the temperature dependence will shed additional light on finite-size effects or the equilibration dynamics at even longer times.

Here we have discussed the finite-temperature real-time properties of magnetic polarons. In a follow-up paper we present our results for the one-hole spectral function in a $t$-$J$ model at finite temperature.

\begin{acknowledgments}
We thank Sebastian Paeckel for helpful discussions. Symmetries in tensor network computations were exploited using the QSpace tensor library ~\cite{Weichselbaum2012,Weichselbaum2020,Weichselbaum2024}. This work was funded in part by the Deutsche Forschungsgemeinschaft under Germany's Excellence Strategy EXC-2111 (Project No.\ 390814868). It is part of the  Munich Quantum Valley, supported by the Bavarian state government with funds from the Hightech Agenda Bayern Plus. F.G. acknowledges funding from the European Research Council (ERC) under the European Union’s Horizon 2020 research and innovation programm (Grant Agreement no 948141) — ERC Starting Grant SimUcQuam. 
\end{acknowledgments}

\appendix

\section{Spin correlations in equilibrium}\label{appendix:spin_correl_in_equil}

\begin{figure}
\centering
\includegraphics[width = 0.47\textwidth]{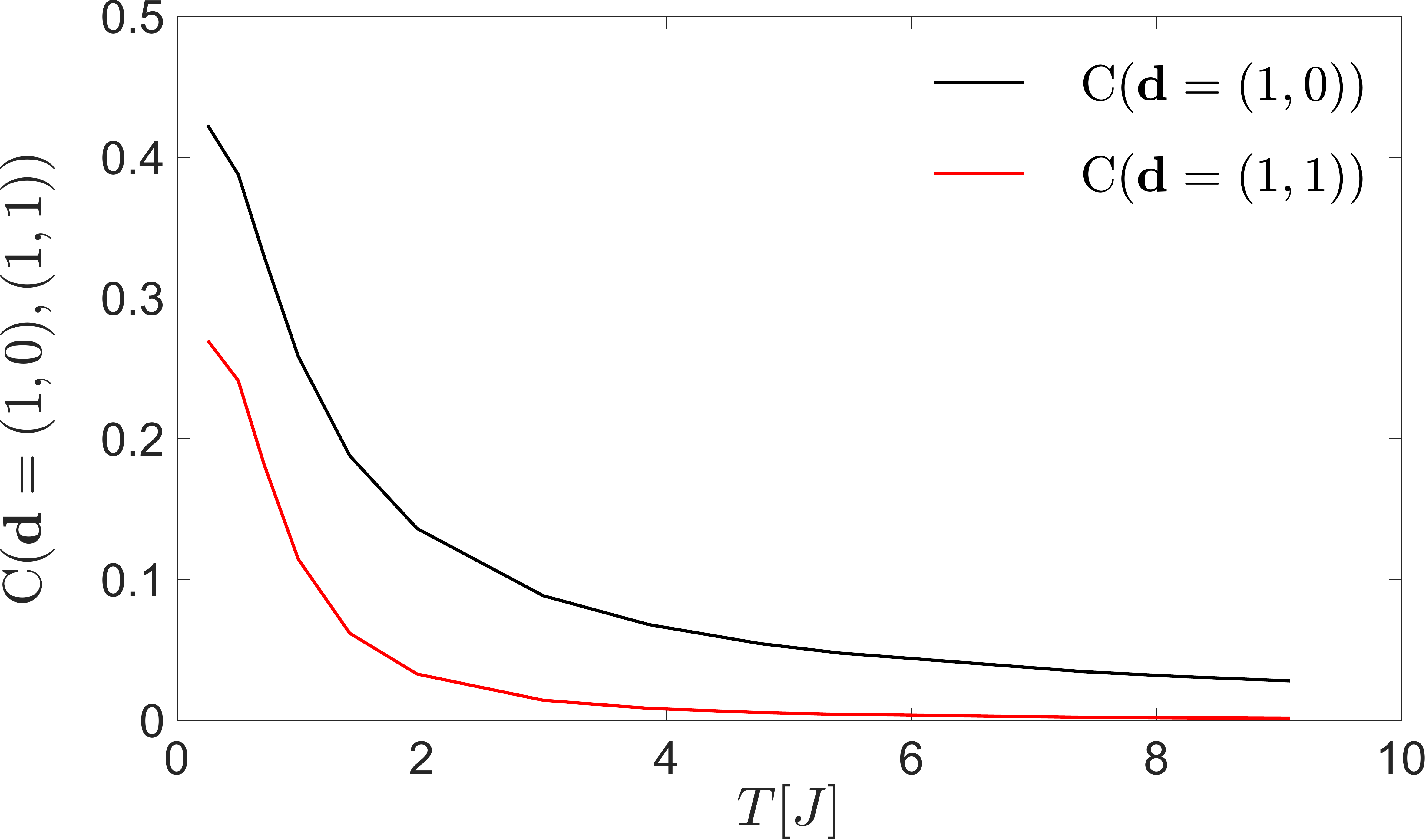}
\caption{Staggered spin correlations $C_{\mathbf{r}} (\mathbf{d})$ (\Eq{eq_sign_corrected_spin_correl}) in equilibrium as a function of temperature. The data presented here was computed on the same $t$-$J$ cylinder as that discussed in the main text and shows nearest (black) and next-nearest-neighbor (red) correlations.}
\label{fig:figure11}

\end{figure}

Given that spin correlations are required for the existence of a finite string tension, it is instructive to observe the presence of spin correlations in equilibrium. In \Fig{fig:figure11} we present the evolution of nearest and next-nearest-neighbor spin correlations in equilibrium as a function of temperature.

\section{Convergence}\label{appendix:convergence}

\begin{figure}
\centering
\includegraphics[width = 0.47\textwidth]{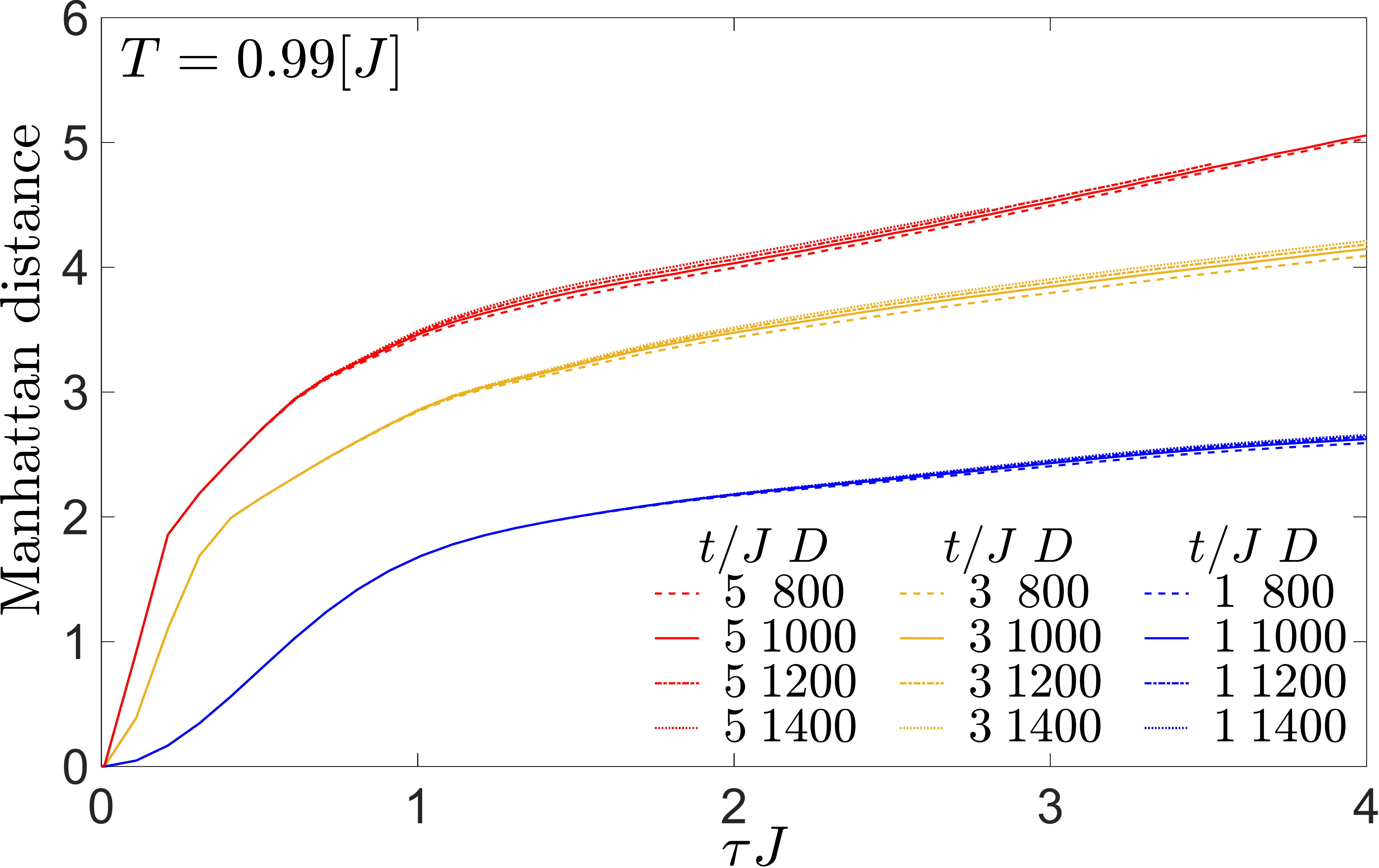}
\caption{Convergence with bond dimension $D$ in the Manhattan distance at $T=0.99J$ for $t/J=1,3,5$.}
\label{fig:figure12}

\end{figure}

The results presented in the main text have been subjected to meticulous analysis with regard to convergence in a number of parameters, including the bond dimension $D$, see \Fig{fig:figure12} for an exemplary analysis.

\begin{figure}
\centering
\includegraphics[width = 0.48\textwidth]{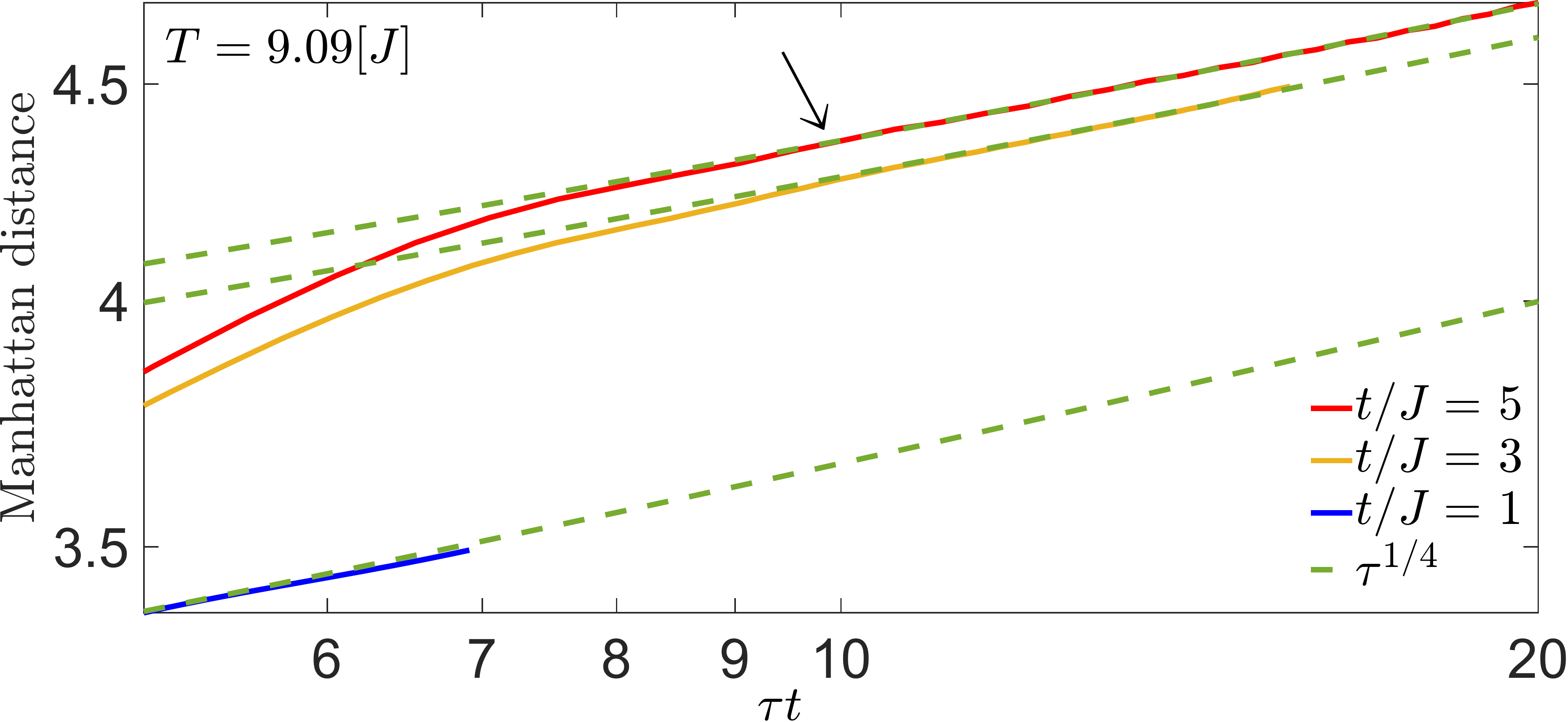}
\caption{Analysis of the high temperature dynamics. Logarithmic plot of the Manhattan distance for different $t/J$ as function of time. The data is plotted against a subdiffusive process (green dashed line), which has been shifted to coincide with the beginning of the respective long-term dynamics. The black arrow indicates the point in time at which the hole has reached the edge of the cylinder.}
\label{fig:figure3}

\end{figure}

\section{Hole spreading at high temperature}\label{appendix:spreading_high_temp}

By approximating the hole motion at infinite temperature as a diffusive quantum random walk on the Bethe lattice with a disorder potential \cite{bohrdt2020dynamical,kanasz2017quantum}, it was shown that the long-term propagation of the hole in the $t$-$J_z$ model is subdiffusive when considering the case of infinite temperature and large $J_{z}/t$. Although our calculations are limited to a finite system size, making it difficult to observe diffusion processes, we have analyzed whether similar subdiffusive behavior can be observed for the more challenging $t$-$J$ model.

In \Fig{fig:figure3} we compare the intermediate to long-time Manhattan distance for our highest temperature $T=9.09J$ against the $\tau^{1/4}$ behaviour (dashed lines) expected for the subdiffusive expansion in the Bethe-lattice model. By scaling the time in units of $1/t$ and displaying the behaviour for different values of $t/J$, we effectively show how the spin coupling $J$ affects the high temperature dynamics. For small spin couplings, i.e. $t/J\gg1$ (red line), we find remarkably good agreement with the subdiffusion process displayed here. It is also worth mentioning that the type of subdiffusion observed here is identical to the subdiffusion previously reported for the $t$-$J_z$ model. 

This subdiffusive behaviour can also be understood in terms of a disorder potential on a Bethe lattice, which slows down the hole expansion. As a result of spin couplings, the movement of the hole from one site to another modifies the energy of the spin system, effectively creating the aforementioned disorder potential. 

Furthermore, it is important to determine the temperature at which the subdiffusion behaviour ceases to exist. As the long-time behaviour for $T \geq 2.99$ remains essentially independent of temperature in the case of $t/J\gg1$, see solid lines in \Fig{fig:figure1}, we conclude that for small spin couplings, i.e., $t/J\gg1$, this subdiffusive behavior persists down to a temperature of $T\approx2J$.

\begin{figure}
\centering
\vspace{1cm}
\subfloat{
\includegraphics[width = 0.48\textwidth]{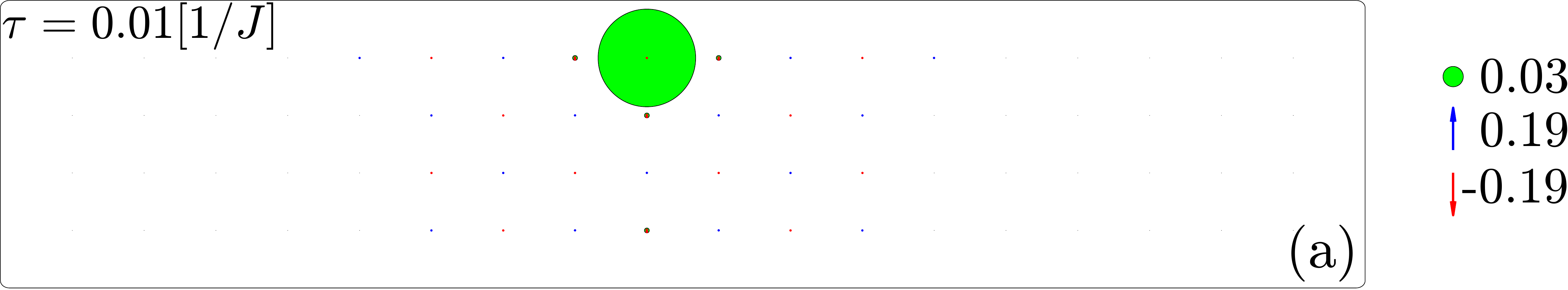}\label{fig:figure12a}}

\vspace{-0.15cm}

\subfloat{
\includegraphics[width = 0.48\textwidth]{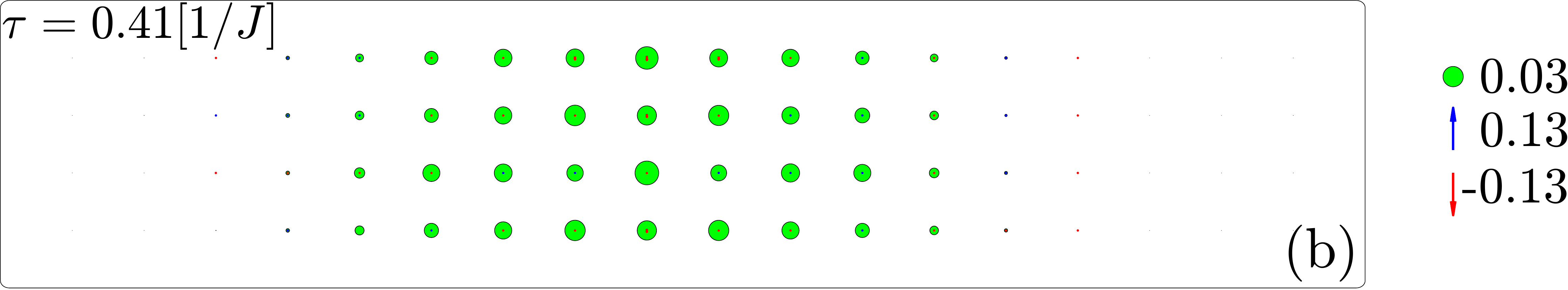}\label{fig:figure12b}}

\vspace{-0.15cm}

\subfloat{
\includegraphics[width = 0.48\textwidth]{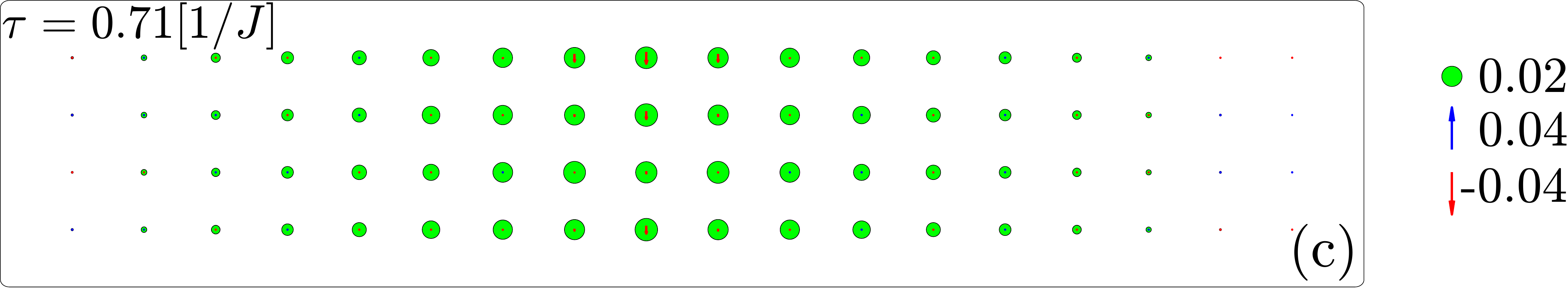}\label{fig:figure12c}}

\vspace{-0.15cm}

\subfloat{
\includegraphics[width = 0.48\textwidth]{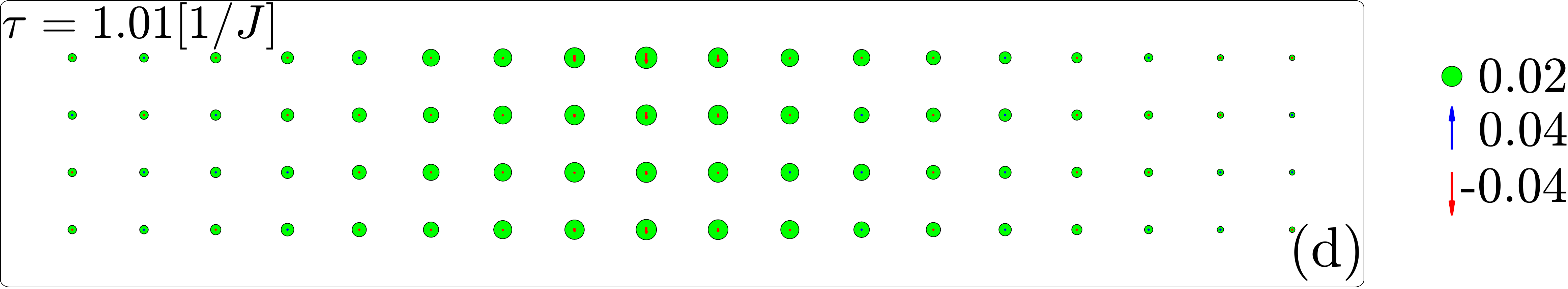}\label{fig:figure12d}}

\vspace{-0.15cm}

\subfloat{
\includegraphics[width = 0.48\textwidth]{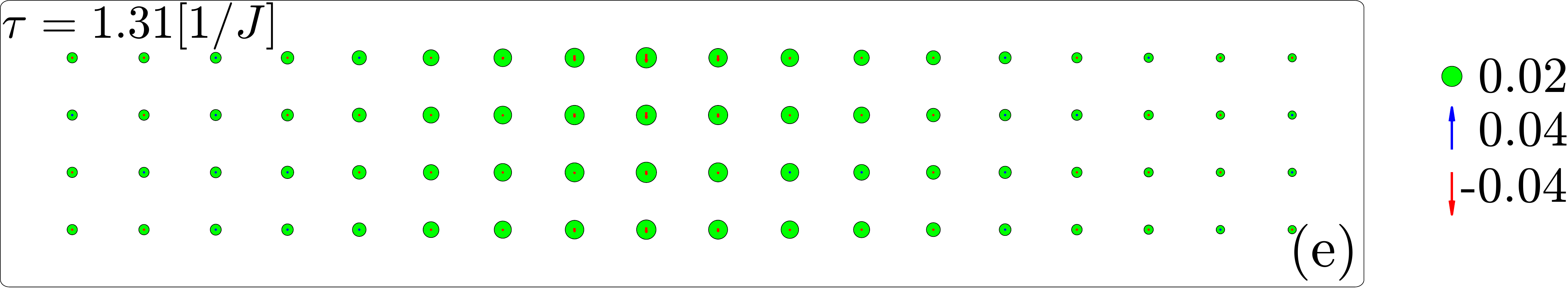}\label{fig:figure12e}}

\vspace{-0.15cm}

\subfloat{
\includegraphics[width = 0.48\textwidth]{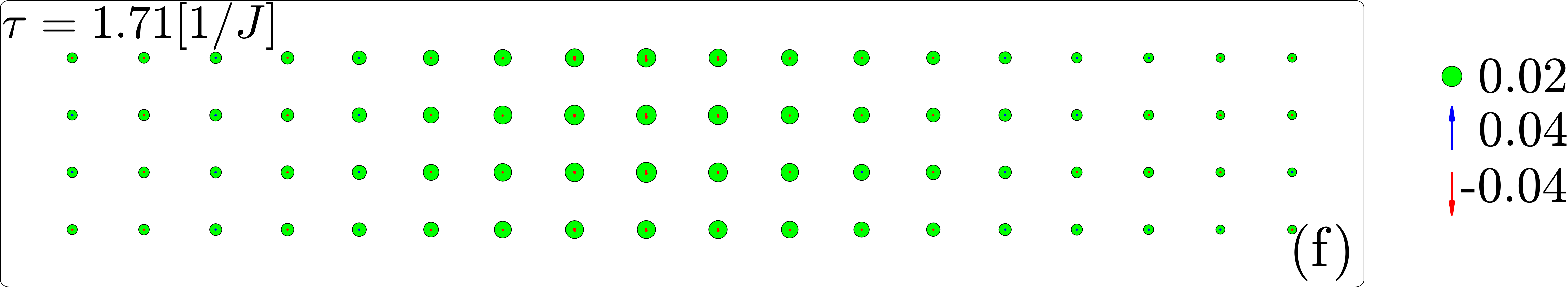}\label{fig:figure12f}}

\vspace{-0.15cm}

\subfloat{
\includegraphics[width = 0.48\textwidth]{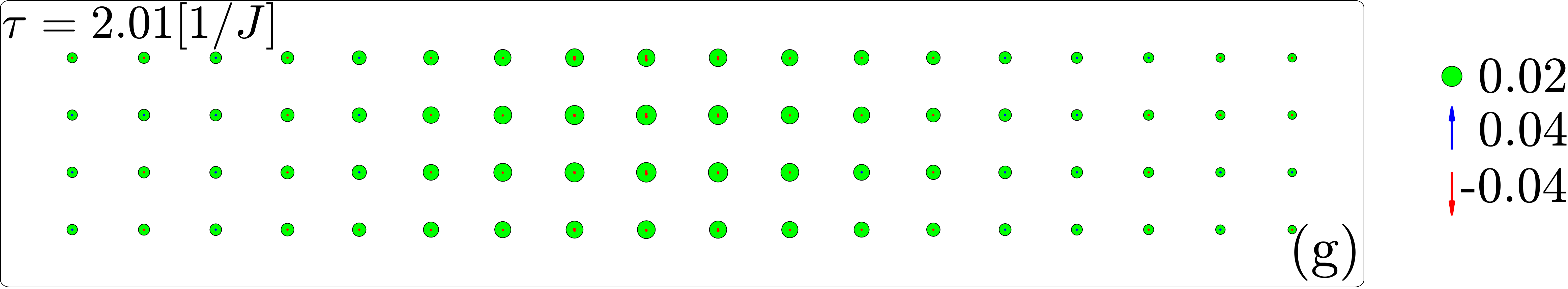}\label{fig:figure12g}}

\caption{Hole density and spin across the entire lattice shown for several times at temperature $T=9.09J$ for $t/J=5$. The data is displayed in the same way as \Fig{fig:figure4}.}
\label{fig:figure_hole_dens_high_temp}

\end{figure}

In order to asses to which extent diffusion can be observed in our finite-size cylinder we extract the point in time at which the hole reaches the left and right edges of the cylinder, see \Fig{fig:figure_hole_dens_high_temp}. If the hole density at all edge sites is above a threshold value $n^h_{\mathrm{tresh}}$, we assume that the hole has reached the left and right edges of the cylinder. We set the threshold value to $n^\mathrm{h}_{\mathrm{tresh}}=0.01$. At $t/J=5$ and $T=9.09J$ we find that the hole reaches the left and right edges of the cylinder at time $\tau \approx 2[1/J]$, which corresponds to the time up to which diffusion can be observed in the system.

\FloatBarrier

\bibliography{paper_bib}

\end{document}